# Continuous-wave and Transient Characteristics of Phosphorene Microwave Transistors

X. Luo[1], K. Xiong[1], J. C. M. Hwang[1], Y. Du[2], and P. D. Ye[2]

[1]Lehigh University, Bethlehem, PA 18015, USA.  [2]Purdue University, West Lafayette, IN 47907, USA.

*Abstract* — Few-layer phosphorene MOSFETs with 0.3-μm-long gate and 15-nm-thick $Al_2O_3$ gate insulator was found to exhibit a forward-current cutoff frequency of 2 GHz and a maximum oscillation frequency of 8 GHz after de-embedding for the parasitic capacitance associated mainly with the relatively large probe pads. The gate lag and drain lag of the transistor was found to be on the order of 1 μs or less, which is consistent with the lack of hysteresis, carrier freeze-out or persistent photoconductivity in DC characteristics. These results confirm that the phosphorene MOSFET can be a viable microwave transistor for both small-signal and large-signal applications.

## I. INTRODUCTION

Since 2013, following the explosive interest in graphene and two-dimensional (2D) atomic-layer transition-metal dichalcogenides (TMDs), there has been a renewed interest in phosphorene as a new 2D material for electronic applications [1]. In less than two years since the invention of phosphorene MOSFET, its small-signal performance in terms of forward-current cutoff frequency $f_T$ and maximum frequency of oscillation $f_{MAX}$ have been pushed above 10 GHz [2]. Meanwhile, the phosphorene MOSFET has been demonstrated to be stable in atmosphere for at least three months and from −50 °C to 150 °C [3]. However, for the phosphorene MOSFET to be a useful microwave transistor, its large-signal transient characteristics must be assessed as shown below.

## II. EXPERIMENT

Similar to [3], the fabrication of phosphorene MOSFETs started with few-layer phosphorene approximately 5-nm thick that was exfoliated from bulk black phosphorus, then transferred onto a high-resistivity silicon substrate coated with 300 nm of $SiO_2$. The transferred phosphorene was then metallized with source and drain electrodes in regions defined by electron-beam lithography to be 1-μm apart. The electrodes consisted of sequentially evaporated Ni and Au layers with thicknesses of 20 nm and 60 nm, respectively. Following source and drain metallization, the phosphorene surface was passivated by $Al_2O_3$ using atomic-layer deposition (ALD). The ALD was seeded by a 0.8-nm-thick pure Al layer, which was expected to getter oxygen from the phosphorene surface as well as the atmosphere and to become completely oxidized before ALD. The ALD was carried out with trimethylaluminum and water precursors at 250 °C until 15 nm of $Al_2O_3$ was deposited. Following surface passivation, a gate electrode was defined by electron-beam lithography in the middle of the source-drain spacing with a length of 0.3 μm and a width of 3 μm. The gate electrode consisted of sequentially evaporated Ti and Au layers with thicknesses of 20 nm and 60 nm, respectively. For parasitics de-embedding, a gateless MOSFET with only interconnects and probe pads was fabricated along with the gated MOSFET.

After the phosphorene MOSFET was fabricated, it was characterized in a Cascade Microtech Summit 1000 thermal probe station under ambient conditions. DC characterization was performed by using an Agilent Technologies 4156C precision semiconductor parameter analyzer. RF characterization was performed by using an Agilent N5230A PNA network analyzer. For gate-lag characterization, an Agilent pulse generator 41501B was used as the gate stimulus while the drain voltage was held constant, and drain current was monitored by an Agilent 6054A digital oscilloscope across a 1-kΩ load resistor in series with the drain supply. For drain-lag characterization, the drain voltage was pulsed while the gate voltage was held constant and the drain current was monitored. For both gate lag and drain lag, the pulse width was 100 μs with a 25% duty cycle.

## III. CONTINUOUS-WAVE CHARACTERISTICS

Fig. 1 shows the output and transfer characteristics of a typical phosphorene MOSFET, which exhibits a *p*-type depletion-mode operation with peak drain current on the order of 100 mA/mm and peak transconductance on the order of 10 mS/mm. Fig. 2 shows both as-measured and de-embedded cutoff frequencies. De-embedding is critical for the small transistor (0.3 μm × 3 μm) with large probe pads (75 μm × 100 μm) so that the intrinsic gate capacitance is approximately 6 fF whereas the parasitic capacitance is approximately 25 fF. Once de-embedded, it can be seen that $f_T$ = 2 GHz and $f_{MAX}$ = 8 GHz, which are higher than that of [3] but lower than that of [2], [2] and [3] being the only reports on the RF performance of phosphorene transistors to date. Compared to [3] with a similar gate insulator of 15-nm $Al_2O_3$ but a gate length of 0.7 μm, the higher cutoff frequencies of the present devices are more than what can be expected from the gate-length reduction alone. The additional improvement is probably from replacing the gallium-arsenide substrate with the silicon substrate, which reduces the interface states and roughness between the transistor channel and the substrate. Compared to

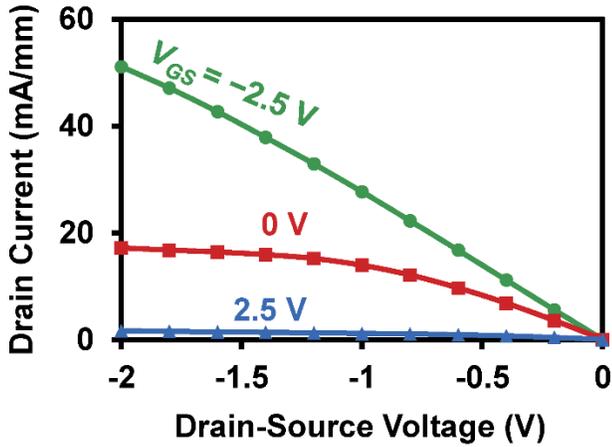

(a)

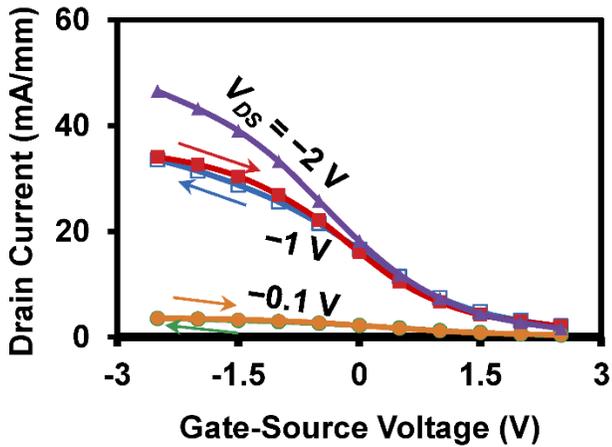

(b)

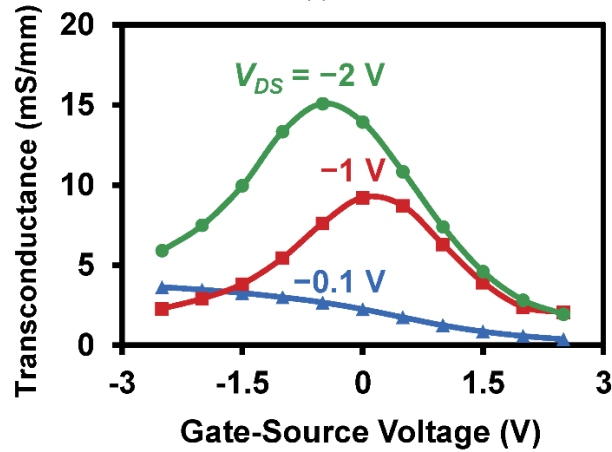

(c)

Fig. 1. (a) Drain characteristics, (b) transfer characteristics, and (c) transconductance of a typical phosphorene MOSFET with 0.3-μm gate length and 15-nm gate insulator of ALD $Al_2O_3$.

[2] with a similar gate length of 0.3 μm but a gate insulator of 21-nm $HfO_2$, the lower cutoff frequencies of the present devices are probably from the lower dielectric constant of $Al_2O_3$ vs. that of $HfO_2$. With the higher dielectric constant of $HfO_2$, dielectric screening of carrier scattering from defect or

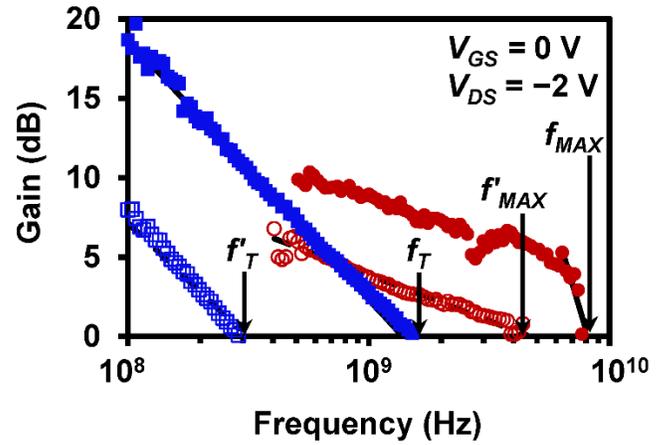

Fig. 2. As-measured cutoff frequencies $f'_T$ and $f'_{MAX}$ and their de-embedded values $f_T$ and $f_{MAX}$ of the phosphorene MOSFET.

impurity can be more effective, resulting in higher carrier mobility and hence, higher transconductance. Nevertheless, all phosphorene MOSFETs, including that of [2] and [3] as well as the present ones exhibit higher $f_{MAX}$ than $f_T$, which is the opposite of most graphene transistors. This implies that phosphorene transistors are more suitable for large-signal microwave operation than graphene transistors.

## IV. TRANSIENT CHARACTERISTICS

Reduction of interface state density, whether between phosphorene and the gate insulator or between phosphorene and the substrate, is crucial to the microwave performance of phosphorene MOSFETs at both small-signal and large-signal levels. The lack of hysteresis in the DC characteristics of Fig. 1(b) implies that there are few interface states of slow response. For faster interface states, gate-lag and drain-lag transient measurements traditionally used on depletion-mode MESFETs or HEMTs can be employed [4], whereas charge-pumping measurements traditionally used on MOSFETs are difficult to implement without a conducting substrate [5].

Fig. 3 shows that the gate and drain lags of the present phosphorene MOSFET are both on the order of 1 μs. Assuming an operating point with −1 V for both the gate and drain voltages, the gate lag was characterized by pulsing the gate voltage from 2.5 V to −1 V while keeping the drain voltage at −1 V, whereas the drain lag was characterized by pulsing the drain voltage from −0.1 V to −1 V while keeping the gate voltage at −1 V. Since the response time of the present pulse test setup is also on the order of 1 μs, the gate and drain lags may be faster than 1 μs, which should be better characterized by using a nanosecond pulsed setup [6]. In any case, the gate and drain lags of the present phosphorene MOSFETs are much better than early GaAs or GaN MESFETs/HEMTs and are consistent with the lack of hysteresis (Fig. 2), carrier freeze out [3], [7], or persistent photoconductivity [8]. The lack of persistent photoconductivity was confirmed by the less than 4% increase of the drain conductance of four phosphorene MOSFETs after 1-min exposure to a 150-W halogen lamp with an intensity on the order of 0.1 W/cm$^2$.

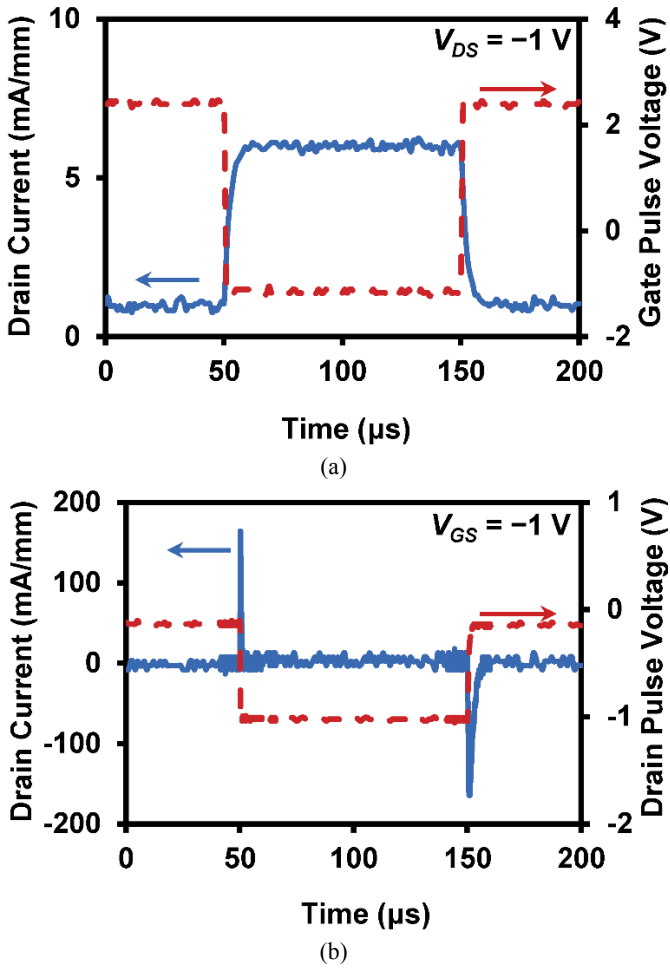

Fig. 3.  (a) Gate-lag and (b) drain-lag transient characteristics of the phosphorene MOSFET.

## V. Conclusion

Despite being a very recent invention, the cutoff frequencies of the present phosphorene MOSFET are already in the microwave range. As expected, further improvement in cutoff frequencies will require not only continued scaling of the gate length, but also reduction of interface states and other parasitics. The transient characteristics of the present phosphorene MOSFET in terms of gate lag and drain lag are on the order of 1 μs or less, which is consistent with the lack of hysteresis, carrier freeze-out or persistent photoconductivity in the DC characteristics. This is very encouraging comparing to that of early GaAs or GaN MESFETs/HEMTs.

## Acknowledgment

This work was supported in part by the U.S. Office of Naval Research under Grant N00014-14-1-0653 and the Air Force Office of Scientific Research and the National Science Foundation EFRI 2-DARE Grant No. 1433459-EFMA.